\documentclass[11pt,aps,prd,superscriptaddress,showpacs,floatfix,nofootinbib]{revtex4}
\usepackage[dvipdfmx]{graphicx}
\usepackage{epsfig}
\usepackage{bm}
\usepackage{amssymb}
\usepackage{wrapfig}
\usepackage{ulem}
\usepackage{multirow}
\usepackage{color}

\newcommand\eL{{\cal L}}

\newcommand\beq{\begin{equation}}
\newcommand\eeq{\end{equation}}
\newcommand\beqa{\begin{eqnarray}}
\newcommand\eeqa{\end{eqnarray}}

\newcommand{\psla}{\mbox{\ooalign{\hfil/\hfil\crcr$p$}}}

\newcommand{\delsla}{\mbox{\ooalign{\hfil/\hfil\crcr{$\partial$}}}}

\newcommand{\vp}{\mbox{$\bm{p}$}}

\newcommand{\vgamma}{\mbox{$\bm{\gamma}$}}
\newcommand{\psibar}{\mbox{${\bar \psi}$}}

\newcommand{\rhSCI}{\mbox{$\rho_{SC}^{({\rm I})}$}}
\newcommand{\rhSCII}{\mbox{$\rho_{SC}^{({\rm II})}$}}

\begin{document}

\title{Spontaneous Spin Polarization due to Tensor Selfenergies 
in Quark Matter}

\author{Tomoyuki~Maruyama}
\affiliation{College of Bioresource Sciences,
Nihon University, Fujisawa 252-8510, Japan}
\affiliation{Advanced Science Research Center, \\
Japan Atomic Energy Agency, Tokai, Naka, Ibaraki 319-1195, Japan}

\author{Toshitaka Tatsumi}
\affiliation{ Department of Physics, Kyoto University, 
Kyoto 606-8502, Japan} 
\affiliation{Advanced Science Research Center, \\
Japan Atomic Energy Agency, Tokai, Naka, Ibaraki 319-1195, Japan}

\date{\today}


\begin{abstract}
We study the magnetic properties of quark matter 
in the NJL model with the tensor interaction.
The spin-polarized phase given by the tensor interaction remains even 
when the quark mass is zero, while the phase given by the axial vector
 interaction disappears.
There are two kinds of spin-polarized phases: one appears 
in the chiral-broken phase, and the other appears in the chiral-restored
 phase where the quark mass is zero.
The latter phase can appear independently of the strength 
of the tensor interaction.
\end{abstract}

\maketitle
 
\newpage
\section{Introduction}

Discovery of magnetars  \cite{mag3,mag}, which are neutron stars with 
super strong magnetic field, seems to revive  an important question 
about the origin of the strong magnetic field in compact stars. 
Magnetars have huge magnetic field of 10$^{15}$G and
are grouped into a new class of compact stars. 
Many people usually assume the conservation of magnetic flux during 
the stellar evolution to explain the magnetic field of the pulsar.
However, if we naively apply this hypothesis to magnetars,
we immediately have a contradiction
that their radius should be much less than the Schwarzschild radius. 
Thus, it may not be very easy to explain the strong magnetic field without
considering properties of hadronic matter inside  stars.
We should pay attention to a microscopic origin 
to solve the ``magnetar'' problem, 

Recently, many theoretical and experimental efforts have been devoted to
explore the QCD phase diagram in the density-temperature plane, which
may be closely related to phenomena observed in relativistic heavy-ion
collisions, compact stars or early universe
\cite{R-ion,R-star,R-star2}. 
In particular, quark-gluon plasma (QGP) at high-temperature but low-density
regime and color superconductivity (CSC) at high-density and
low-temperature regime have been elaborately studied
\cite{R-ion,csc}. 

Because dense matter occupies a large portion of compact stars,
its property should be reflected in various phenomena.
In Ref.~\cite{tat00}  one of the author (T.T.) has suggested a possibility
of a ferromagnetic transition in QCD; 
it is possible in quark matter interacting with one-gluon-exchange
interaction and its critical density is order of nuclear density,
$\rho_{FM} \simeq \rho_0$, where $\rho_0$ is normal nuclear matter density.  
Using this idea we can roughly estimate the strength of the magnetic
field at the surface of compact stars. 
Considering a star with mass,
$M\sim 1.4 M_{\odot}$,  and radius, $R\sim 10$Km, and assuming the
dipole magnetic field, the maximum strength at the surface can be 
simply estimated by 
$B_{\rm max}=(8\pi/3) f_Q\mu_q \rho_0$, where $f_Q$ is the volume fraction
of quark matter and $\mu_q$ the quark magnetic moment. 
Thus we evaluate it as $O(10^{15-17}{\rm G})$ for the extreme case, 
$f_Q=1$, which should be compared with observations. 
This is a perturbative result based on the Bloch mechanism, 
in analogy with electron gas \cite{blo29,her66,mor85}. 

In the relativistic framework the ``spin density'' can take 
the two forms  \cite{MaruTatsu}, 
$\psi^\dagger \Sigma^i \psi (\equiv - \psibar \gamma_5 \gamma^i \psi )$
and $\psi^\dagger \gamma^0 \Sigma^i \psi (\equiv - \psibar \sigma^{12} \psi )$,
with $\psi$ being the quark field.
The former is a space-component of the axial-vector (AV) mean-field, and the
later is that of the tensor (T) one.
These two mean-fields become equivalent to each other in 
the non-relativistic limit, while they are quite different in 
the ultra-relativistic limit (massless limit) \cite{MaruTatsu}.
In the text we shall call  the former and latter polarization  the
AV-type and  T-type spin polarizations (SP), respectively.     

For quark matter, we have introduced the AV interaction
and have studied the SP mechanism 
in the mean-field approximation \cite{tat00,NT05,MNT}.
In theses studied we have succeeded to show the co-existence of 
the spin polarization and the color super-conductivity (CSC) 
\cite{nak03A} and the dual chiral density wave (DCDW) \cite{NT05,YT15}.

Furthermore, Maedan have also studied the SP in the NJL model
with this AV mean-field  \cite{maedan07}.
When the mass is fixed, the SP
appears in high density region, but the AV mean-field disappears 
when the quark mass becomes zero.
Then, the spin-polarized phase can  appear in small density region 
just lower than the chiral phase transition density. 

As mentioned above, the AV channel of two quark interaction has often been used
for the SP study in quark matter because
this channel is obtained by the Fierz transformation 
from the one-gluon exchange (OGE) interaction.
On the other hand, the T channel has not been often used first
because this interaction channel does not appear in the Fierz transformation 
from the OGE interaction.
However, low-energy effective QCD models such as the NJL model have
not been constructed based on the OGE interaction, and 
there is not any reason to exclude this channel.

The T channel interaction can play an important role differently from 
the AV channel interaction to produce the spin-polarized phase because
the T-type SP can appear 
even if the quark mass becomes zero \cite{MNT}.
Actually, Tsue et al. \cite{TPPY} has also shown that 
the SP appears in the chiral-restored phase, where the quark mass is zero, 
in the NJL model within the effective potential approach.
   
In addition, the magnetic interaction of quark matter 
with the T-type SP is much larger than that
of the AV-type SP \cite{MNT}.
In the Fermi degenerate system, the magnetic field should be almost created
by magnetization, which is proportional to 
$\psi^\dagger \gamma^0 \Sigma^i \psi$.
The lower component of the Dirac spinor contributes to  
$\psi^\dagger \Sigma^i \psi$ and
$\psi^\dagger \gamma^0 \Sigma^i \psi$, oppositely.
In the relativistic region, where the quark mass is much less than the Fermi
momentum, the contribution from the lower component has the same order
of that from the upper component. 
As  $<\psi^\dagger \Sigma^i \psi>$ increases, then, 
$<\psi^\dagger \gamma^0 \Sigma^i \psi>$ becomes smaller 
in the AV-type SP.

Thus, the AV-type SP appears in narrow density region
below the chiral transition 
and may not contribute to the magnetic field very largely.
In contrast, the T-type SP can appear in the wide density 
region and largely contribute to the magnetic field.
Thus, we should examine behaviors of the SP and its
relation with chiral symmetry.

In this paper we study the T-type SP in the NJL
model and figure out the relation between 
the spontaneous SP and chiral transition.
In the next section we present a framework to deal with the present subject.
In Sec. 3 we show the results of the numerical calculation and discuss
the relation between SP and chiral restoration.
Sec. 4 is devoted to summary and concluding remarks.   

\newpage

\section{Formalism}
%

\subsection{Lagrangian and Quark Propagator}

In order to examine the T-type SP
 we start with the following NJL-type Lagrangian density 
with  $SU(2)$ chiral symmetry,
\begin{equation}
\eL = \eL_{K} + \eL_s + \eL_V + \eL_T
\end{equation}
with
\begin{eqnarray}
\eL_{K} &=& \psibar (i \delsla - m) \psi ,
\\ 
\eL_S &=& - \frac{G_s}{2} \left[ (\psibar \psi )^2 
+ (i \psibar \gamma_5 \tau \psi )^2 \right] ,
\label{Schn}
\\ 
\eL_V &=& - \frac{G_v}{2} \left[ 
(\psibar \gamma_\mu \psi )(\psibar \gamma^\mu \psi )
+ (i \psibar \gamma_5 \gamma_\mu \tau \psi )  
(i \psibar \gamma^\mu \gamma_5 \tau \psi )
\right] ,
\label{Vecn}
 \\
\eL_T &=&
- \frac{G_T}{2} \left[ 
(\psibar \sigma_{\mu \nu} \psi ) (\psibar \sigma^{\mu \nu} \psi ) 
+ (\psibar i \tau_a \gamma_5 \sigma_{\mu \nu} \psi ) 
(\psibar i \tau_a \sigma^{\mu \nu} \gamma_5 \psi ) \right] ,
\end{eqnarray}
where $\psi$ is a field operator of quark, 
$G_s$, $G_v$ and $G_T$ are the coupling constants for the scalar, vector 
and tensor channels, respectively.

Here, we comment on the tensor interaction.
If the original Lagrangian includes only $\eL_S$ 
in Eq.~(\ref{Schn}), 
the Fierz transformation effectively gives the following Lagrangian:
\begin{eqnarray}
\eL_{FT} & =&
 \frac{1}{4} G_s \left[ 
 \left( \psibar \psi \right)^2 + \left( \psibar i \tau \gamma_5 \psi \right)^2
-  \left( \psibar \tau \psi \right)^2
-  \left( \psibar i \gamma_5  \psi \right)^2
+ 2 \left( \psibar \gamma_5 \gamma_\mu \psi \right) 
\left( \psibar \gamma_5 \gamma^\mu \psi \right)
\right.
\nonumber \\ 
&& \left.
- 2 \left( \psibar \gamma_\mu \psi \right) 
\left( \psibar \gamma^\mu  \psi  \right)
+ \frac{1}{2} \left( \psibar \sigma_{\mu \nu} \psi \right)
\left( \psibar \sigma^{\mu \nu} \psi \right)
- \frac{1}{2} \left( \psibar \sigma_{\mu \nu} \tau \psi \right)
\left( \psibar \sigma^{\mu \nu} \tau \psi \right)
 \right] .
\end{eqnarray}
Thus, the T channel of the interaction can appear even if the
original interaction does not include this channel.

In the present work, we restrict calculations and discussions 
to the flavor symmetric matter ($\rho_u=\rho_d$) at zero temperature.
Within the mean-field approximation the quark Dirac spinor 
$u({\vp,s})$ is obtained as the solution of the following equation,
\begin{equation}
\left[ \psla - M_q - U_0 \gamma^0 -U_T \Sigma_z \right] u(\vp,s) = 0
\label{Dirac-UT}
\end{equation}
with $\Sigma_z= {\rm diag}(1,-1,1,-1)$ and 
\begin{eqnarray}
M_q &=& - G_s \rho_s = - G_s < \psibar \psi> ,
\label{ScEq}
\\
U_0 &=& G_v \rho_q =  G_v < \psibar\gamma^0 \psi> ,
\label{VcEq}
\\
U_T &=&  G_T \rho_T  = G_T <\psibar \Sigma_z \psi>
- G_T <\psibar \Sigma_z \tau_3 \psi> \tau_3 .
\label{TsEq}
\end{eqnarray}

In the mean-field approximation  the quark Green
function is defined as a solution of the following equation:
\begin{equation}
\left[ \psla - M_q - U_0 \gamma^0 - U_T \Sigma_z \right] S(p) = 1.
\label{qGreen}
\end{equation}
By solving the above Eq.~(\ref{qGreen}) we can obtain
\begin{equation}
S(p) = \frac{\left[ \gamma_\mu p^{* \mu} + M_q + \Sigma_z U_T \right] 
\left\{ p^{* 2} - M_q^2 + U_T^2 
+ 2 U_T  ( p_z \gamma_5 \gamma^0 - p_0 \gamma_5 \gamma^3  ) \right\} }
{(p_0^{* 2} - E_p^2 - U_T^2)^2 - 4 U_T^2 (\vp_T^2 + M_q^2) \pm i \delta }
\end{equation}
with $p_\mu^* = p_\mu - U_0 \delta^0_\mu$ and  
$E_p = \sqrt{\vp^2 + M_q^2}$.

The $S(p)$ has poles at $p_0 = \pm e(\vp,s)$,
which give single particle energies
\begin{equation}
e(\vp,s) = \sqrt{(\sqrt{M_q^2+\vp_T^2} + s U_T )^2 + p_z^2} + U_0
 = \sqrt{ E_p^2 + 2 s U_T \sqrt{M_q^2+\vp_T^2} + U_T^2 } + U_0 ,
\label{enSDT}
\end{equation}
where $s =\pm 1$ indicates the spin of a quark.

It should be interesting to compare it with a single particle energy 
in the AV mean-field, $U_A$:
\begin{equation}
e(\vp,s) = \sqrt{(\sqrt{M_q^2+ p_z^2} + s U_A)^2 + \vp_T^2} + U_0
 = \sqrt{ E_p^2 + 2 s U_A \sqrt{M_q^2+p_z^2} + U_A^2 } + U_0 .
\label{enSDA}
\end{equation}

Here, we make a comment on the difference in the SP
between the tensor and axial-vector interactions.
When $U_A=U_T$, we can obtain the above expression of  $e(\vp,s)$ 
in Eq.~(\ref{enSDA}) from that in Eq.~(\ref{enSDT})  
by exchanging $p_z$ and $p_T$.
The surfaces in the momentum space at the fixed energy
have the same relation between  the two types of SP.
However, $p_z$ is one-dimensional while $p_T$ is the absolute value
of the two dimensional vector.

\begin{figure}[htb]
\vspace*{-2em}
\hspace*{-2cm}
{\includegraphics[scale=0.55,angle=270]{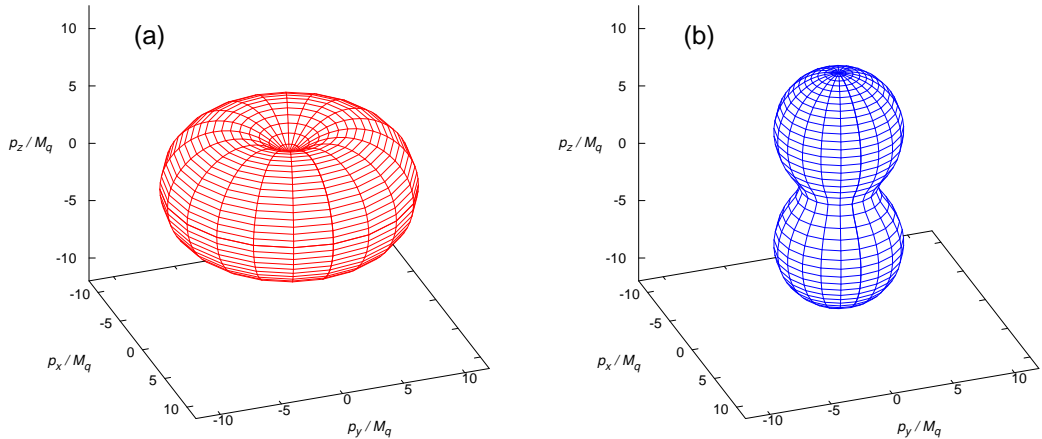}}
\vspace*{-5em}
\caption{The energy constant surfaces for $e - U_0 = 3M_q$ and $s=-1$,
when $U_T = 3M_q$ (a) and when $U_A = 3M_q$ (b).}
\label{MomDis}
\end{figure}

In Fig.~\ref{MomDis}, we show the constant energy surface 
for $e(\vp,-1) - U_0 = 3 M_q$
in the T-Type spin-polarized phase when $U_T=3M_q$ (a) 
and in the AV-type spin-polarized phase when $U_A=3M_q$ (b).
We see that difference in the momentum distribution between the two types of
the SP:
it is deformed prolately in the T-type SP and oblately in the AV-Type SP.

Using these single particle energies, the quark propagator is
separated into the vacuum part $S_F$ and the density dependent part
$S_D$  as
\begin{equation}
S(p) = S_F (p) + S_D (p). 
\end{equation}
with
\begin{eqnarray}
 S_F (p) &=&  \frac{\left[ \gamma^\mu p^*_\mu + M_q + \Sigma_z U_T \right] 
\left\{ p^2 - M_q^2 + U_T^2 
+ 2 U_T  ( p_z \gamma_5 \gamma^0 -  p_0 \gamma_5 \gamma^3 ) \right\} }
{ \left[ p_0^2 -e^{*2}(\vp,1) + i \delta \right]
\left[ p_0^2 -e^{*2}(\vp,-1) + i \delta \right] } ,
\\
S_D(p) &=& \sum_{s=\pm 1}
\left[ \gamma_0 e^* - \vgamma \cdot  \vp + M_q + \Sigma_z U_T \right] 
\left\{1 + \frac{ s  ( p_z \gamma_5 \gamma^0 -  p_0 \gamma_5 \gamma^3 ) + s U_T}
{ \sqrt{\vp_T^2 + M_q^2}}\right\} 
\nonumber \\ && \quad \quad  \times
 \frac{i \pi}{2 e^*(\vp,s)} n(\vp,s) \delta[p_0 - e(\vp,s)] ,
\end{eqnarray}
where $p^*_\mu - p_\mu - U_0 \delta^0_\mu$, $e^* = e - U_0$, 
$n(\vp,s) = \Theta[e_F - e(\vp,s)]$, and $e_F$ is the Fermi energy.

In the above expression, the quark density is written as
\begin{eqnarray}
\rho_q  &=& N_d \sum_{s=\pm 1} 
\int \frac{d^3 p}{(2\pi)^3} \theta[ e_F - e(\vp,s) ] ,
\label{qRh}
\end{eqnarray}
where $N_d = N_f N_c = 6$ is given by the degeneracy of the flavors,
$N_f = 2$, and color degrees of freedom, $N_c = 3$;
 the baryon density is given as $\rho_B = \rho_q/N_c$.

Here, we give a comment about the vector channel of the interaction. 
The vector mean-field, $U_0$, has only a role to shift the
single particle energy and does not influence a spin property.
Without $U_0$, the quark chemical potential, $e_F$, does not monotonously 
increase as  density becomes larger. 
The system transits from the density with the maximum chemical potential
to the chiral restoration; this transition is the first order phase transition. 
As the $G_v$ increases, the transition density becomes larger,
and, when the vector field is sufficiently large, the phase transition is
of the second order\footnote{If the phase-transition is of the first
order, the chiral phase transition occurs when the dynamical mass is finite.
The density dependences of the dynamical mass in the density region
below the chiral phase transition are the same 
independently of the order of the phase transition. } \cite{TMF91,MTF92}.

Thus, we can control the phase transition with the vector interaction
without changing magnetic properties.
We assume that the vector coupling $G_v$ is large, and that 
the chiral transition is of the second order.
We rewrite $p_0^*$ and $e^*$ to $p_0$ and $e$  and 
eliminate $U_0$ in the following.

In the mean-field approximation
the dynamical quark mass $M_q$ and the $U_T$ are determined by
\begin{eqnarray}
1 - \frac{G_s \rho_s}{M_q} &=&0 ,
\label{NJLEqS}
\\
1 - \frac{G_T \rho_T}{U_T} &=&0 ,
\label{NJLEqT}
\end{eqnarray}
where the scalar density $\rho_s$ and the tensor density are given by
\begin{eqnarray}
\rho_s &=& N_d \int \frac{d^4 p}{(2 \pi)^4} {\rm Tr} \left[ i S(p) \right] ,
\\
\rho_T &=& N_d \int \frac{d^4 p}{(2 \pi)^4} {\rm Tr} 
\left[ i \Sigma_z S(p) \right] .
\end{eqnarray}

The scalar density is separated into two parts, the vacuum part and the
density-dependent part as
\begin{eqnarray}
&& \rho_s = \rho_s(V) + \rho_s(D) = 
N_d \int \frac{d^4 p}{(2 \pi)^4}  {\rm Tr} \left[i S_F(p) \right]
+ N_d \int \frac{d^4 p}{(2 \pi)^4}  {\rm Tr} \left[i S_D(p) \right] .
\end{eqnarray}
The  density dependent part is written as 
\begin{eqnarray}
\rho_s (D) &=& N_d
\sum_{s=\pm 1} \int \frac{d^3 p}{(2 \pi)^3}  ~n(\vp,s)~
\frac{M_q}{e(\vp,s)} \left(1  + \frac{s U_T} {\sqrt{M_q^2+p_T^2}} \right) .
\label{RhSD}
\end{eqnarray}

The density dependent part of the tensor density is also written as
\begin{eqnarray}
\rho_T(D) &=&
N_d  \sum_{s=\pm 1} \int \frac{d^3 p}{(2\pi)^3} ~ n(\vp,s)~
\frac{ s \sqrt{\vp_T^2 + M_q^2} + U_T }{e(\vp,s)} .
\label{RHST}
\end{eqnarray}
Because $\rho_T <0$ when $U_T >0$,  so that Eq.~(\ref{NJLEqT})
has a solution when $G_T <0$.

This argument is right only when the SP is isoscalar,
where the average spin of  u- and d-quarks are directed 
along the same direction.
In the isovector spin-polarized system the tensor densities 
for u- and d-quark have opposite signs.
In the symmetric matter, we define 
$\rho_T = \rho_T(u) - \rho_T(d) = 2\rho_T(u)$ and 
$U_T = U_T(u) - U_T(d) = 2U_T(u)$, and  rewrite Eq.~(\ref{TsEq}) as 
$U_T = - G_T \rho_T$, which is the same as  Eq.~(\ref{NJLEqT})
except the sign of r.h.s.
This fact demonstrates that, when $G_T>0$, the isovector spin-polarized
phase can appear,
and its strength is the same as that for $G_T<0$.

Here, we give a comment on the tensor density.
When $M_q=0$, the tensor density in Eq.~(\ref{RHST}) becomes
\begin{equation}
\rho_T(D) = - \frac{N_d}{12 \pi} e_F^3  \neq 0 ,
\end{equation}
while $\rho_A = 0$ \cite{NT05,nak03A}.
When $M_q = 0$, namely, the T-type SP can appear 
while the AV-type SP never appears.
This difference comes from the momentum distribution
in the SP phase (see Fig.~\ref{MomDis}).

In this paper we per form the argument only when $G_T <0$, 
but we can apply the same argument 
to the isovector spin-polarized system for $G_T>0$; 
the latter system has a larger magnetization because of the opposite sign
for the u- and d-quark charges.

In order to extract the vacuum part we use 
the proper time regularization (PTR) \cite{SchPTR51}, where 
 the thermodynamical potential density is written as 
\begin{eqnarray}
\Omega_{vac}
&=& i N_d \int  \frac{d^4 p}{(2 \pi)^4} 
\ln \left[ (p_0^2 - e^2(\vp,+1)) \right] \left[ (p_0^2 - e^2(\vp,-1)) \right]
\nonumber \\
&=& -i N_d \sum_{s = \pm 1}
\int_0^{\infty} \frac{d \tau}{\tau}  \int  \frac{d^4 p}{(2 \pi)^4} 
e^{\tau \left[ (p_0^2 - e^2({\bf p },s)) \right] }
\nonumber \\
& \approx & \frac{N_d}{8 \pi^2} \sum_s \int_{1/\Lambda^2}^{\infty} 
\frac{d \tau}{\tau^2} \int_{M_q}^{\infty} d E_T E_T 
e^{- \tau (E_T + s U_T)^2 }
\end{eqnarray}
at zero temperature, where $\Lambda$ is the cut-off parameter.
The vacuum part of the  scalar density is then given by 
\begin{equation}
\rho_s (V)  = \frac{ \partial \Omega_{vac} }{ \partial M_q~~} =
- \frac{N_d M_q}{8 \pi^2} \sum_s \int_{1/\Lambda^2}^{\infty} 
\frac{d \tau}{\tau^2} e^{- \tau (M_q + s U_T)^2 } 
=  - \frac{N_d M_q}{8 \pi^2} \Lambda^2 \sum_s
F_2 \left( \frac{(M_q + s U_T)^2}{\Lambda^2} \right) ,
\end{equation}
where the function $F_n$ is defined as
\begin{eqnarray}
&& F_n(x) = x \int_{x}^{\infty} \frac{d \tau}{\tau^n} e^{- \tau } .
\end{eqnarray}

The vacuum part of the tensor density can be also obtained with  
$\rho_T(V) = \partial \Omega_{vac} / \partial U_T$.
However, this term strongly
depends on the cut-off parameter $\Lambda$ in the present model,
even when $M_q = 0$.
Indeed, the vacuum part of the tensor density is also dependent on
the regularization scheme and the value of the cut-off parameter.
(see Appendix.\ref{VacAmb} for details).
For example, the vacuum contribution in the PTR suppresses the SP
while that in the effective potential approach \cite{TPPY}
enlarges it.

In the AV type SP, the vacuum part in the PTR also
suppresses the SP \cite{NT05}, 
while that in the momentum cut-off enlarges it \cite{maedan07}.

The cut-off parameter has a role to restrict the momentum space 
in the calculation.  
The T and AV densities are given by the difference 
between the spin-up and spin-down contributions, which depends 
on the restriction of the momentum space, so that
the result sensitively depends on the regularization method and the value 
of the cut-off parameter.

In the usual renormalization procedure we regularize the vacuum
polarization by introducing the suitable counter terms which are determined 
from physical values.
In order to regularize the tensor density, $\rho_T(V)$ 
we need to introduce at least
three counter terms which are proportional to $U_T^2$,  $U_T^4$ and
$U_T^2 M^2$.
In the NJL model we regularize the vacuum contributions 
by using a cut-off parameter.
The vacuum part of the scalar density is associated with the dynamical
quark mass in the vacuum, but not concerned with the spin properties.
In the present model, the vacuum part of the tensor density strongly 
diverges as $\Lambda \rightarrow \infty$ for small asymmetry of
the spin states.
This fact does not have any physical meaning, but we do not have
any clear rule to regularize the tensor density in a systematic way. 

Thus, the cut-off dependence of the tensor density from the vacuum
contribution is less meaningful at present.
In the NJL model it is not easy to apply a consistent method 
even for the qualitative discussions.
In the next section, then, we perform actual calculation without 
the vacuum contribution for the tensor density: $\rho_T \approx \rho_T(D)$.

In Appendix \ref{SPwV}, instead, we try to give a temporal calculation
for the SP including it.

\newpage

\section{Results}
\label{SecRes}

In this section we show some numerical results for SP
in relation with chiral restoration.
For this purpose,
we consider the chiral limit and use two kinds of the parameter-sets 
PM1 ($G_s \Lambda^2 = 6, \Lambda = 850$ MeV)
and  PM2 ($G_s \Lambda^2 = 6.35, \Lambda = 660.37$ MeV)
from Ref.~\cite{NT05}.

\begin{wrapfigure}{r}{8.3cm}
\begin{center}
\vspace{-0.5cm}
{\includegraphics[scale=0.47,angle=270]{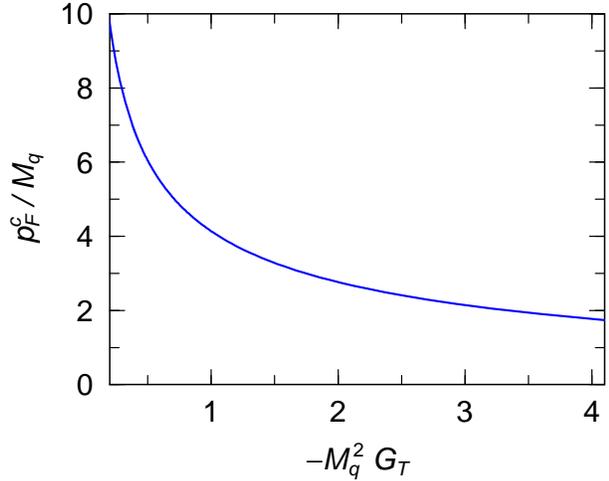}}
\caption{\small Coupling constant versus the critical Fermi momentum 
for the spontaneous spin polarization. \label{CrRHOT}}
\end{center}
\end{wrapfigure}

Before showing actual numerical results, we discuss the critical
density of the SP ($\rho_{SC}$).

The tensor mean-field is determined by the following self-consistent equation
\footnote{$\rho_T \rightarrow 0$ when $U_T \rightarrow 0$, Eq.~(\ref{RHST}).}:
\begin{equation}
F_T (U_T) =1 - \frac{G_T \rho_T}{U_T} = 0 .
\label{UTeq}
\end{equation}

When $U_T \gg e_F+ M_q$, $\rho_T = -N_d e_F^3 /12 \pi$ and 
$\rho_q = N_d U_T e_F^2/4 \pi$  (see Sec.~\ref{AsecRh}), 
so that $e_F \rightarrow 0$ and  $F_T (U_T) \rightarrow 1$ 
when $U_T \rightarrow \infty$ at the fixed density.
Hence, Eq.~(\ref{UTeq})  has a solution when $F_T(0) < 0$ at $U_T = 0$
which leads to
\begin{eqnarray}
&& J =  1 + \frac{G_T N_d}{2 \pi^2} \left\{  p_F E_F
+ \frac{M_q^2}{2} \ln \left(\frac{E_F + p_F}{E_F - p_F} \right) \right\} 
\le 0 , \quad
\label{FTcond}
\end{eqnarray}
where $p_F$ is the Fermi momentum, and 
$E_F = e_F(U_T=0) = \sqrt{p_F^2 + M_q^2}$.

$J=0$ in Eq.(\ref{FTcond}) can be expressed by the two independent parameters,
$M_q^2 G_T$ and $p_F^c/M_q$, where $p_F^c$ 
indicates the critical Fermi momentum for the spontaneous SP.
We show the boundary of the spin-polarized phase in Fig.~\ref{CrRHOT}.

\begin{wrapfigure}{r}{8.7cm}
\begin{center}
{\includegraphics[scale=0.47]{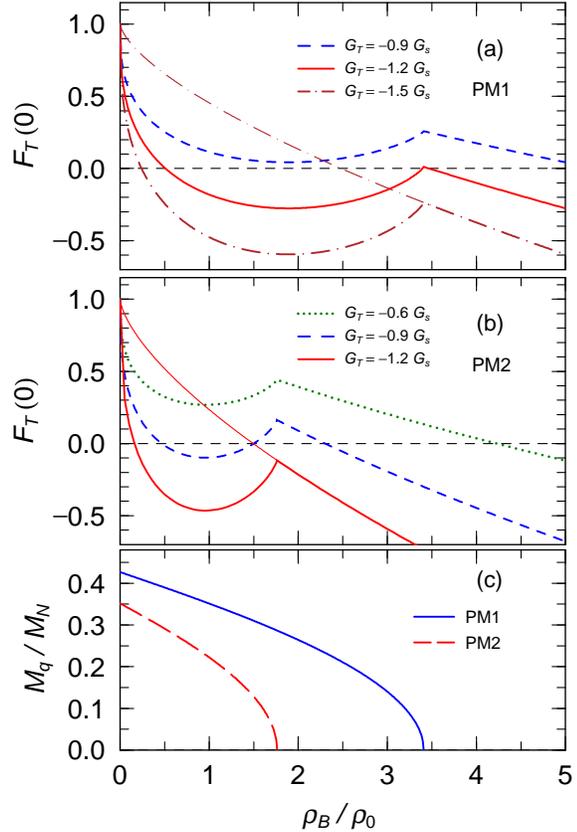}}
\caption{\small (Color online) $F_T(0)$ with  PM1 (a) and PM2 (b) and  
 the dynamical quark mass (c) as functions of $\rho_B / \rho_0$.
In the upper (a) and middle (b) panels the dotted, dashed, solid and 
dot-dashed lines represent the results  when $G_T/G_s = -0.6$, $-1.2$
and  $-1.5$, respectively.
The thin lines in the upper (a) and middle panels (b) 
indicates $F_T(0)$ when  $M_q=0$.
In the lower panel (c) the solid and long dashed lines indicate 
$M_q/M_N$ with PM1 and PM2, respectively. 
\label{FTfig} }
\end{center}
\end{wrapfigure}

In Fig.~\ref{FTfig} we show  $F_T(0)$ with PM1 (a) and PM2 (b) 
when $- G_T/G_s = 0.6 \sim - 1.5$ 
as functions of baryon density, $\rho_B$, normalized by
the normal nuclear density $\rho_0 = 0.17$fm$^{-3}$. 
In addition, we show  the dynamical quark mass normalized by
nucleon mass $M_N$ at the spin-saturated system ($\rho_T=0$) 
with PM1 (solid line) and PM2 (long dashed line). 

As baryon density becomes larger, $F_T(0)$ decreases at first, 
and increases later.
so that  $F_T(0)$ has a maximum at the chiral phase transition (CPT) 
density, $\rho_c$, and monotonously decreases, again.

It was shown in Ref.~\cite{MNT} that $F_T(0)$ monotonously decreases
when the dynamical quark mass $M_q$ is fixed.
However, the dynamical quark mass is also decreasing in the NJL model,
and this effect enlarges $F_T$ up to $\rho_B = \rho_c$, 
where the quark dynamical mass becomes zero.

For comparison, we show $F_T (0)$ when $M_q=0$  with thin lines,
where we plot the results only when $G_T = -1.5 G_s$ for PM1 and
 $G_T = -1.2 G_s$ for PM2. 
We see that $F_T(0)$ monotonously decreases when $M_q=0$ 
with the increase of $\rho_B$.
Because $\partial M_q/\partial \rho_B$ is not continuous, 
the maximum point of $F_T(0)$ becomes a cusp, which corresponds to that
the CPT is of the second order\footnote{If the phase transition is 
of the first order, the critical density of the chiral phase transition 
is lower that that of the second order.}.

As mentioned before, $F_T(0) =0$ shows the critical density of the phase
transition between the spin-saturated and spin-polarized phases.
As the coupling $-G_T$ becomes larger, the number of the crossing points
turns to be one, three and one.
The last case, where the number of the crossing point is one, 
indicates that with $F_T(0) < 0$ at $\rho_B = \rho_c$.
In this case the line of $F_T(0)$ with $M_q =0$ also crosses zero at the
density lower than the CPT density 
(see thin lines in Fig.~\ref{FTfig}).

These results suggest that there are two kinds of the spin-polarized phases: 
one is  the chiral-broken SP (SP-I) which appears in the chiral-broken
phase, $M_q >0$, and the other is the chiral-restored SP (SP-II) which
appears in the chiral-restored phase, $M_q=0$.
Here we define $\rhSCI$ and $\rhSCII$ as critical
densities of the SP-I and SP-II phases, respectively.
$\rhSCII$ exits with any $G_T$, but $\rhSCI$  appears only
when  $-G_T$, is large, namely the SP-II phase always appears 
independently of the strength of the tensor interaction.
In addition, ${-G_T}$ becomes further larger, $F_T(0) < 0$ at $\rho_B = \rho_c$ 
$\rhSCII < \rho_c$;
here, we should note that $M_q=0$ is a solution of Eq.~(\ref{ScEq}) as well as 
a solution of the gap equation (\ref{NJLEqS}).

\begin{figure}[thb]
\begin{center}
\vspace{0.5cm}
{\includegraphics[scale=0.55,angle=270]{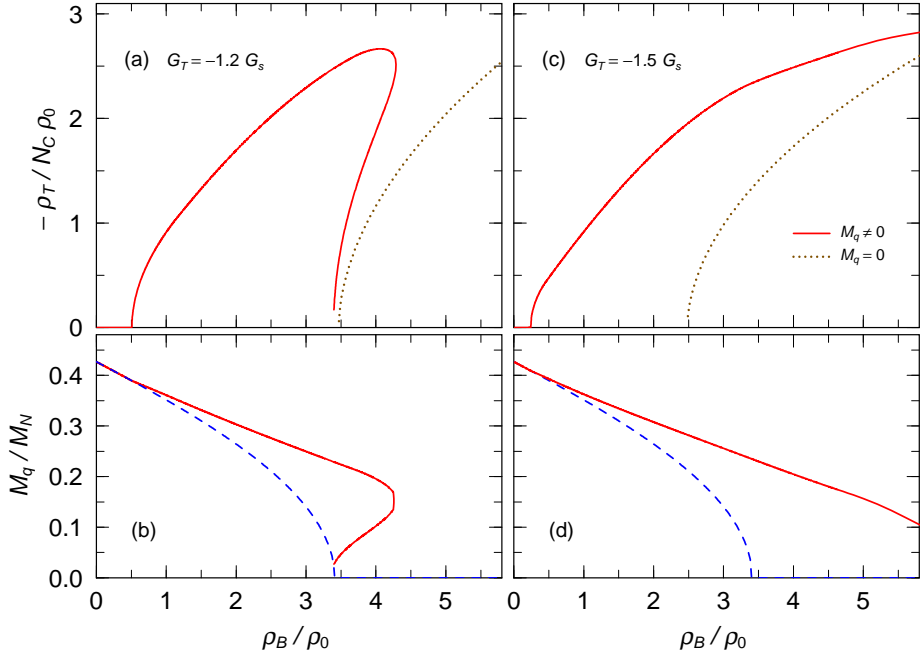}}
\caption{\small 
Spin polarization properties with PM1.
Upper panels (a,c) : the tensor densities normalized 
by the normal nuclear density.
The solid and dotted lines represent the results in the chiral symmetry
 broken and restored phases, respectively. 
Lower panels (b,d): the dynamical quark mass normalized by nucleon mass 
in the spin-polarized (solid lines) and spin-saturated phases (dashed line).
The left and right panes show the results 
when $G_T = -1.2$ (a, b), $=-1.5$ (c, d), respectively.}
\label{TSPM1}
\end{center}
\end{figure}

In Fig.~\ref{TSPM1}, we show the baryon density dependence 
of the tensor density $\rho_T/\rho_0$ with PM1 (upper panel) 
and that of the dynamical mass (lower panel) when
$G_T = - 1.2 G_s$ (a,b) and $G_T = -1.5 G_s$ (c,d).
In the upper panel the solid lines represent $\rho_T/(N_c \rho_0)$ 
in the spin-polarized phase when $M_q > 0$, and the dotted lines indicates that 
when $M_q=0$. 
In the lower panel, the solid and dashed lines represent 
the dynamical quark mass in the spin-polarized and spin-saturated
phases, respectively. 

When $G_T = -1.2 G_s$ (a,b),
we can see that two kinds of spin-polarized phases, SP-I and SP-II
phases, appear.
In addition, there are density region where the three solutions 
corresponding to  the spin-polarized phases, 
two SP-I phases and one SP-II phase, exist.

When $G_T = -1.5 G_s$ (c,d), 
the SP-I phase appears at first, and 
 the SP-II phase appears in  the density region,  $\rho_B < \rho_c$.
Both the two SP phases exist in a same density region up to a density
larger than the CPT density, $\rho_c$, and the SP-I phase
disappears at a density larger than $\rho_c$, where $M_q = 0$ and
$U_T \neq 0$.

In this approach we discard the vacuum contribution to the tensor density
though the scalar density includes the vacuum part,
so that we cannot define the total energy and cannot determine 
what is realized among the spin-saturated, SP-I and SP-II phases.

\begin{wrapfigure}{r}{8.8cm}
\begin{center}
\vspace{-0.2cm}
{\includegraphics[scale=0.43,angle=270]{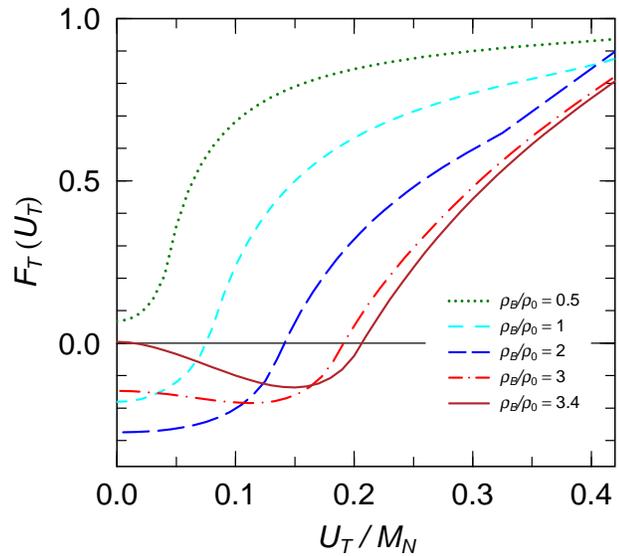}}
\caption{\small (Color online) $F_T(U_T))$ versus $U_T/M_N$ with PM1 and
 $G_T = - 1.2 G_s$.
The dotted, dashed long-dashed, dot-dashed and solid lines represent
results when $\rho_B/\rho_0 = 0.5, 1, 2, 3, 3.4$, respectively. }
\label{FT-R}
\end{center}
\end{wrapfigure}

In order to look into this behavior more clearly, we calculate 
$F_T(U_T)$ by varying baryon density.
In Fig.~\ref{FT-R} we show the results at several  baryon densities.

When $\rho_B \lesssim 2 \rho_0$, $F_T$ is a monotonously increasing function.
As the density decreases,
$F_T(0)$ becomes smaller, and, when $F_T(0) < 0$, 
the equation $F_T(U_T) =0$ has a solution.

When $\rho_B \gtrsim 2\rho_0$, $F_T(0)$ becomes larger with the increase
 of the density, but $F_T(U_T)$ has a minimum at a certain $U_T$.
The equation $F_T(U_T) =0$ has two solutions when $F_T(0) >0$, and the
 minimum value is negative.    
As density further increases, the minimum value of $F_T$ becomes
 positive, and there is no solution.

\begin{figure}[bht]
\begin{center}
\vspace{0.5cm}
{\includegraphics[scale=0.55,angle=270]{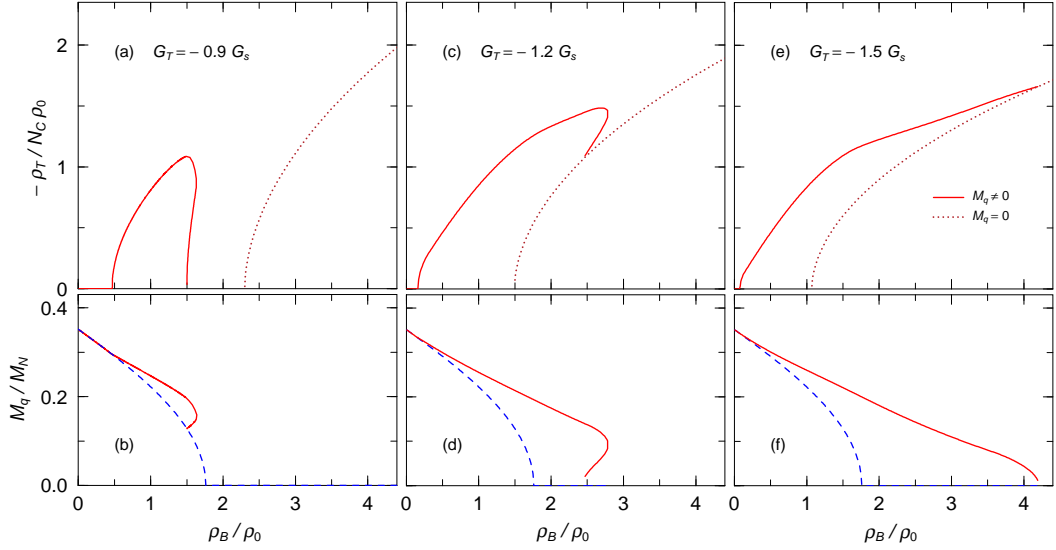}}
\caption{\small Spin polarization properties with PM2.
 Upper panels (a,c,e): the T densities normalized 
by the normal nuclear matter density.
The solid and dotted lines represent the results in the chiral 
 broken and restored phases, respectively. 
Lower panels (b,d,e): the dynamical quark mass normalized by nucleon
mass in the spin-polarized (solid lines) and spin-saturated phase (dashed
 line).
The left, middle and right panes show the results 
  when $G_T/G_s = -0.9$ (a,b), $G_T/G_s =-1.2$ (c,d), 
$G_T/G_s = -1.5$ (e,f), respectively. }
 \label{TSPM2}
\end{center}
\end{figure}

In Fig.~\ref{TSPM2}, we show the results with PM2 
with $G_T= -0.8 G_s$ (a,b), $G_T= -1.2 G_s$ (c,d) and $G_T= -1.5G_s$ (e,f).
The behaviors of the SP are similar to those in PM1 
(Fig.~\ref{TSPM1}).
When $G_T= -0.8 G_s$ (a,b), the SP-I and SP-II phases appear in different
density region.
When $G_T= -1.2 G_s$ (c,d), and $G_T= -1.5 G_s$ (e,f),
the results clearly show that
the SP-I and SP-II phases simultaneously exist in the density region,
$\rho_B > \rho_c$, and that 
the SP-I phase disappears at higher density.

\begin{figure}[hbt]
\begin{center}
{\includegraphics[scale=0.6]{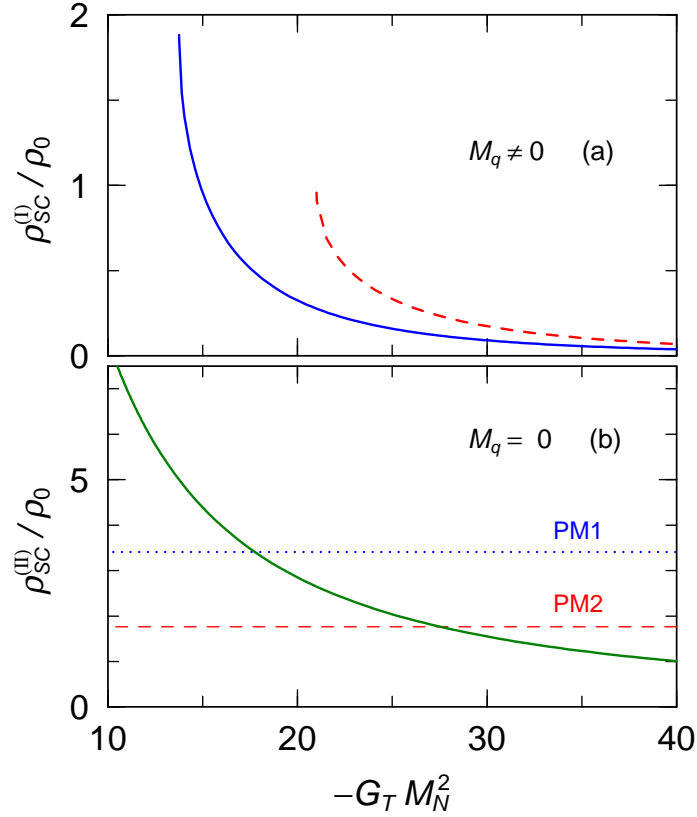}}
\caption{\small (Color Online) Critical density between the spin-saturated
 and spin-polarized phases as functions of $G_T M_N^2$ 
when $M_q \neq 0$ (a) and $M_q = 0$ (b).
In the upper panel the solid and dashed lines represent 
$\rhSCI/\rho_0$ with
PM1 and PM2, respectively. \label{SpCrRh} 
In the lower panel the solid line shows the critical density of SP-II, 
$\rhSCII/\rho_0$, and
the dotted and dashed lines indicate the critical densities of CPT, 
$\rho_c/\rho_0$, with PM1 and PM2, respectively.}
\end{center}
\end{figure}

We can confirm that the SP-I phase disappears at a density larger 
than $\rho_c$, where the tensor density is finite, $\rho_T \neq 0$.
In Fig.~\ref{SpCrRh} we finally show the critical density 
between the spin-saturated and spin-polarized phases as a function 
of $G_T M^2_N$ in the chiral-broken phase, $\rhSCI$, (a) 
and chiral-restored phase, $\rhSCII$, (b).
The critical density when $M_q=0$, $\rhSCII$, is determined 
only by $G_T$, independently of $G_s$. 
In addition, we plot the critical density of CPT, $\rho_c$, with PM1
(dotted line) and PM2 (dashed line) in Fig.~\ref{SpCrRh}b.

We see the results in the chiral phase as follows.
As $-G_T$ increases, the phase transition in the SP-I phase 
appears at $\rho_B \approx 1.9 \rho_0$  when  $- G_T M_N^2 = 13.8$ 
($-G_T/G_s = 0.949$) in PM1
and at $\rho_B \approx 0.96 \rho_0$ when  $- G_T M_N^2 = 21.0$ 
($-G_T/G_s = 0.819$) in PM2.
In the chiral-restored phase  $\rhSCII \rightarrow \infty$ as 
$-G_T \rightarrow 0$, so that the phase transition occurs at any value of
$-G_T$.
As $-G_T$ increases, the critical density of SP-II, $\rhSCII$, 
becomes smaller, and then it is lower than the CPT density,
$\rhSCII < \rho_c$, when $- G_T M^2 > 17.7$ ($G_T/G_s < - 1.21$) 
in PM1 and when $- G_T M^2 > 27.6$ ($G_T/G_s < - 1.07$) in PM2.

\newpage

\section{Summary}

We have studied the spontaneous SP of quark matter in the NJL model 
with the tensor interaction.
There appear two kinds of the spin-polarized phases, the SP-I and SP-II phases, 
where the dynamical quark mass is non-zero and zero, respectively.
The SP-I phase appears when the T coupling $G_T$ is negatively large, 
but the SP-II phase can always appears above the critical density
when $G_T < 0$ though its transition density depends on $G_T$ \cite{MNT,TPPY}.

The SP-I phase can exist in the density region above the CPT density 
and shifts the chiral transition to higher density.
On the other hand, the SP-II phase can appear below the CPT density.
The SP-I and SP-II phases can exist at the same density when $-G_T$ is large.
In the present model we cannot discuss the stability of each phase and make
a critical conclusion.
However, we can easily suppose that the phase transition between
the SP-I and SP-II phases is of the first order.

We have considered an appearance of a non-uniform phase with the AV
interaction during the chiral transition, where pseudo-scalar condensate 
as well as scalar condensate is non-vanishing, called 
as dual chiral density wave (DCDW) \cite{NT05}. 
The T-type SP must leads to a new type DCDW, which can appear in the chiral
restored phase.
We should study it in future.

In this paper we have made the discussion only when $G_T <0$: 
the spin-polarized phase is isoscalar. 
When $G_T > 0$, the spin-polarized phase becomes isovector where the
directions of the SP for $u$ and $d$ quarks are opposite.
The strength of the magnetic field is much larger 
in the isovector spin-polarized phase
than in the isoscalar SP phase because the charge of $u$ and $d$ quarks
have opposite signs.
 
In the present work we have discarded the vacuum contribution 
to the tensor density  because its contribution strongly depends 
on the regularization method.
We have demonstrated in Appendix~\ref{VacC} that the vacuum
contribution becomes important at high densities.  
However, the value of the cut-off parameter is determined to reproduce the
dynamical quark mass in the vacuum, but is not related to the spin property,
and then the large dependence on the cut-off parameter is not meaningful.

In order to remove the ambiguity we need to use a renormalizable model and
to introduce counter terms to reproduce the vacuum spin-susceptibility 
at zero temperature,
which is determined with the other model such as the lattice QCD.
It is a future problem.

In this work, furthermore, we have not considered the AV channel of quark-quark
interaction, which can be derived by the Fierz transformation of the
one gluon exchange.
The calculation of the spin-polarized phase is very difficult 
when both the T and AV interactions are introduced 
because the momentum distribution is very complicated.  
If the weak T interaction is mixed with the AV one, however, 
the AV-type SP phase appears even when the quark mass is zero; 
we have not discuss it in this paper.

When the quark mass is small, the tensor density decrease as the
AV-type SP becomes larger 
\footnote{$\rho_T \rightarrow 0$ when $U_A \rightarrow \infty$.}, 
and the magnetic field 
which is produced by the spin-current also decreases \cite{MNT}.
However, the magnetic field can be kept to be finite  
by the tensor mean-field even if it is weak.
So, the mixing of AV and T interactions  may exhibit a new
spin-polarization in quark matter.

In future, we hope to develop our formulation in the system including
both the AV and T interactions. 

\acknowledgments
This work was supported in part by the Grants-in-Aid for the Scientific
Research from the Ministry of Education, Science and Culture of
Japan~(16k05360).
T.T. is partially supported by Grant-in-Aid for Scientific Research on Innovative Areas through No. 24105008 provided by MEXT.

\newpage

\appendix

\section{Density Dependent Parts of Densities}
\label{AsecRh}

In this section, we give the detailed expressions of the quark density
$\rho_q$ , the scalar density $\rho_s$ and
 the tensor density $\rho_T$ 
with the quark mass $M_q$, the chemical potential $e_F$ and
the T field  $U_T (>0)$.

\subsection{Quark Densities}

When $U_T < e_F - s M_q$ for $s=\pm 1$,
\begin{eqnarray}
\rho_q (s)  &=&  \frac{N_d}{2 \pi^2} \left\{ 
\frac{1}{6} \sqrt{e_F^2 - (M_q+sU_T)^2 }
\left[ 2 e_F^2- (M_q+sU_T)(2M_q-sU_T) \right] 
\right. \nonumber \\ && \left. \quad \quad 
- \frac{s}{2} U_T e_F^2 \left[ \frac{\pi}{2}
 - \sin^{-1} \left( \frac{M_q+sU_T}{e_F} \right) \right] \right\} ,
\label{qRh-A2}
\end{eqnarray}


When $U_T > e_F - s M_q$,
\begin{equation}
\rho_q (+1) = 0 , \quad 
\rho_q (-1)  =    \frac{N_d}{4 \pi} U_T  e_F^2 .
\label{qRh-A2b}
\end{equation}

\bigskip

\subsection{Scalar Densities}

When $U_T < e_F - s M_q$ for $s=\pm 1$,
\begin{eqnarray}
\rho_s (s) &=& 
\frac{N_d}{4 \pi^2} M_q  \left[ e_F \sqrt{e_F^2 -(M_q+sU_T)^2} \right.
\nonumber \\ && \quad   \quad   \quad  
\left. - \frac{(M_q+sU_T)^2}{2} \ln \left( 
\frac{e_F + \sqrt{e_F^2 -(M_q+sU_T)^2}}{e_F - \sqrt{e_F^2 -(M_q+sU_T)^2}} \right)
\right] ,
\label{RhS-A2}
\end{eqnarray}


When $U_T > e_F - s M_q$,
\begin{equation}
 \rho_s (\pm 1) = 0 ,
\end{equation}

\bigskip

\subsection{Tensor Density}

When $U_T < e_F - M_q$ for $s=1$ or $U_T < M_q$ for $s=-1$ 
\begin{eqnarray}
\rho_T (s) &=& \frac{N_d}{12 \pi^2} s \left\{ 
(M_q+sU_T)^2 \left( -  M_q + \frac{s}{2}U_T \right)
\ln \left( \frac{e_F + \sqrt{e_F^2 -(M_q+sU_T)^2}}
{e_F - \sqrt{e_F^2 -(M_q+sU_T)^2}} \right)
\right. \nonumber \\ && \left. \quad\quad\quad\quad
+ ~ e_F (M_q-2sU_T) \sqrt{e_F^2 -(M_q+sU_T)^2}
\right. \nonumber \\ && \left. \quad\quad\quad\quad
+ e_F^3 \sin^{-1} \left(\frac{\sqrt{e_F^2 -(U_T + sM_q)^2}}{e_F}\right) \right\} .
\label{TRh-A2}
\end{eqnarray}
When $U_T > e_F -M_q$ for $s=1$,
\begin{equation}
\rho_T(+1) = 0 .
\end{equation}

When $e_F + M_q > U_T > M_q$ for $s=-1$,
\begin{eqnarray}
\rho_T (-1) &=& - \frac{N_d}{12 \pi^2} \left\{
- \frac{1}{2} (U_T - M_q)^2 \left( U_T + 2M_q \right)
\ln \left( \frac{e_F + \sqrt{e_F^2 -(U_T - M_q)^2}}
{e_F - \sqrt{e_F^2 -(U_T - M_q)^2}} \right)
\right. \nonumber \\ &&  \quad\quad\quad\quad
+ ~ e_F \left( 2U_T + M_q \right) \sqrt{e_F^2 -(U_T - M_q)^2}
\nonumber \\ && \left .\quad\quad\quad\quad
+ ~ e_F^3 \left[ \pi -  \sin^{-1} 
\left(\frac{\sqrt{e_F^2 -(U_T-M_q)^2}}{e_F} \right) \right]  \right\} .
\label{TRh-A2a}
\end{eqnarray}
When $U_T > e_F + M_q$  for $s=-1$, 
\begin{eqnarray}
\rho_T (-1) &=&  - \frac{N_d}{12 \pi}  e_F^3 .
\label{TRh-A2b}
\end{eqnarray}

\newpage

\section{Vacuum Contribution to Tensor-Field}
\label{VacC}

\subsection{Ambiguities of Vacuum Contribution}
\label{VacAmb}

In the text we mentioned that the vacuum contribution is
ambiguous and dependent on the regularization scheme.
We  explain the reason of this difference with the energy
cut-off and the three-dimensional momentum cut-off as examples.

In these regularization schemes, the vacuum part of the tensor density 
is written as
\begin{eqnarray}
&& \rho_T(V) = N_d \int \frac{d^4 p}{(2 \pi)^4} {\rm Tr} 
\left[ i \Sigma_z S_F(p) \right] .
\nonumber \\
 &=& i N_d \sum_s \int \frac{d^4 p}{(2 \pi)^4} n_V(\vp,s)
\frac{ 4s U_T(- p_0^2 - M^2 - U_T^2 - \vp_T^2 + p_z^2 ) }
{(-2e(\vp,s))[e^2(\vp,1) - e(\vp,-1)^2] [p_0 + e_s(\vp,s) - i \delta]}   ,
\nonumber \\
 &=& - N_d  \sum_{s=\pm 1} \int \frac{d^3 p}{(2\pi)^3} ~ n_V(\vp,s)~
\frac{ s \sqrt{\vp_T^2 + M_q^2} + U_T }{e(\vp,s)} ,
\label{RHTS-VR}
\end{eqnarray}
where $n_V(\vp,s)$ is an effective momentum distribution for negative
energy particles including the cut-off parameter.

In the energy cut-off regularization, we should take 
$n_V = \Theta[\Lambda_e - e(\vp,s)]$;
apparently the $\rho_T(V)$ is the different sign of $\rho_T(D)$;
in the present choice $\rho_T(D) < 0 < \rho_T(V)$.
Namely, the vacuum contribution suppresses the tensor density. 

In general the cut-off parameter is taken to be much larger than the
Fermi energy, $\Lambda_e \gg E_F$, and the total tensor density becomes
positive, $\rho_T(V) + \rho_T(D)>0$, so that the SP does not appear 
when $G_T<0$.
When $G_T>0$, however, the spontaneous SP appears in the vacuum 
when the cut-off, $\Lambda_e$, increases and exceeds 
a certain critical value; this phenomenon is not though to have
any physical meaning.

In the momentum cut-off regularization, on the other hand, 
the effective momentum
distribution is taken to be $n_V =\Theta(\Lambda_p - |\vp|)$.
When $0< U_T \ll 1$,
\begin{eqnarray}
\rho_T(V) &\approx& 
-N_d  \sum_{s=\pm 1} \int \frac{d^3 p}{(2\pi)^3} ~ \Theta(\Lambda_p - |\vp|)~
\frac{ s \sqrt{\vp_T^2 + M_q^2} + U_T }{E_p} 
\left( 1 - \frac{s U_T \sqrt{\vp_T^2 + M_q^2}}{E_p^2} \right) ,
\nonumber \\
 &\approx& 
-2 N_d U_T \int \frac{d^3 p}{(2\pi)^3} ~ \Theta(\Lambda_p - |\vp|)~
 \frac{p_z^2}{E_p^3} ~<~ 0.
\label{RHTS-MC}
\end{eqnarray}
$\rho_T(V)$ has the same sign of $\rho_T(D)$, namely
the vacuum contribution enlarges the tensor density. 

The vacuum contribution to the tensor density is determined by
the two effects: one is the difference in the volume in the momentum space 
between the spin-up  the spin-down quarks, 
and the other is momentum dependence of
 $\sum_s {\bar u}(\vp,s) \sigma_{12} u (\vp,s)$ at the fixed momentum.
Two effects have opposite roles; the former reduces
the tensor density, and the latter increase it.

In the energy cut-off regularization, the former effect is larger, 
and the vacuum contribution reduces the SP.
In the momentum cut-off regularization, in contrast, 
the former effect does not exist,
and then the vacuum contribution increases the SP.

When $\Lambda_{e,p} \gg M_q$ and $U_T \ll 1$, the vacuum contribution becomes
$\rho_T(V) \approx N_d U_T \Lambda_e^2 /2 \pi^2$ for the energy cut-off 
and $\rho_T(V) \approx - N_d U_T \Lambda_p^2 /3 \pi^2$ 
for the momentum cut-off .
Both results are proportional to the square of the cut-off parameter
though the signs of the two results are opposite.

\newpage

\subsection{Spin Polarization with Vacuum Contribution}
\label{SPwV}

In this section, we discuss the vacuum polarization in the proper time 
regularization. 
  
The vacuum contribution is given by
\begin{equation}
\rho_T (V) =  \frac{N_d}{4 \pi^2} \Lambda^2 
\int_{M-U_T}^{M+U_T} d E_T F_1 \left( \frac{E_T^2}{\Lambda^2} \right) 
+  \frac{N_d}{8 \pi^2} U_T \Lambda^2 \sum_s 
F_2\left[\frac{(M + sU_T)^2}{\Lambda^2}\right]  .
\end{equation}

In the limit of $\Lambda \rightarrow \infty$, the tensor density becomes
\begin{eqnarray}
\rho_T (V)   
 &\approx& 
 \frac{N_d}{4 \pi^2}  \left\{  \Lambda^2 U_T +
 (M_q^2 U_T -  \frac{1}{3} U_T^3) \ln \frac{\Lambda^2}{|M_q^2-U_T^2|}
  \right.
\nonumber \\
&& \left. \quad
- \frac{1}{3} M_q^3 \ln \left( \frac{M_q+U_T}{M_q - U_T} \right)^2
+  \frac{1}{3} M_q^2 U_T  - \frac{5}{9} U_T^3 
 \right\} 
\label{RhTL}
\end{eqnarray}
When $U_T \ll M_q$, in addition, 
\begin{eqnarray}
\rho_T (V) &\approx&
 \frac{N_d}{4 \pi^2}  \left\{
\Lambda^2 U_T + (M_q^2 U_T -  \frac{1}{3} U_T^3) \ln \frac{\Lambda^2}{M_q^2}
-  M_q^2 U_T  - \frac{4}{9} U_T^3 \right\}  
\label{RhTLM}
\end{eqnarray} 
Thus, the terms proportional to $U_T$, $U_T^3$ and $M_q^2 U_T$ diverge
in the limit of $\Lambda \rightarrow \infty$.

The above equation shows us that the spin-susceptibility proportional
to $\partial \rho_T /\partial U_T$  has a very
large value at any density, 
and the SP does not appear in the chiral-broken phase.

When $M_q = 0$ and $U_T \ll 1$, it becomes 
$\rho_T(V) \approx N_d U_T \Lambda^2 / 4 \pi^2$ 
 and 
$\rho_T(D) \approx -N_d U_T p_F^2 / 2 \pi^2$, and the condition of the
SP becomes 
\begin{eqnarray}
p_F^2 \ge \frac{\Lambda^2}{2} - \frac{2 \pi^2}{G_T},
\end{eqnarray}
which is strongly dependent on the cut-off parameter $\Lambda$.

In the field theory, the divergent terms are renormalized to be
physical values.
In the NJL model, the cut-off parameter is taken to be finite,
and  determined by the quark mass at zero density.
On the other hand we cannot relate the vacuum part of the tensor density  
with any physical quantity.
In addition, even the sign of this part depends on the regularization method.
we cannot believe such a large contribution from the vacuum.

In the present model the term proportional to $\Lambda^2$ makes
 $\rho_T(V)$ extraordinarily large.
In the AV-type SP phase  \cite{NT05}, on the other hand, 
the vacuum contribution of the AV density under the AV-field, $U_A$, 
is written, when $|U_A| \ll M_q \ll  \Lambda$, as
\begin{eqnarray}
\rho_A (V)  &\approx&
\frac{N_d}{\pi^{2}} U_A M_q^2 \ln \left( \frac{\Lambda}{M_q} \right) .
\end{eqnarray}
We see that this vacuum contribution in the AV-type SP \cite{NT05} 
does not affect the final result as largely as that of the T-type SP.

As shown in the energy cut-off calculations in Appendix~\ref{VacAmb},
the term proportional to $\Lambda^2$ indicates a contribution 
from the surface area of the integration in region
restricted with the cut-off parameter in the momentum
space, and it must be removed by the renormalization.
 
So, we examine the vacuum contribution  by removing 
the term proportional to $\Lambda^2$. 
For this purpose we introduce an additional counter term, $\beta_T$ and 
define the renormalized thermodynamical potential density as 
\begin{equation}
\Omega_R = \Omega_{vac} -  \frac{1}{2} \beta_T U_T^2,
\label{OmR1}
\end{equation}
which gives the renormalization vacuum tensor density as
$\rho_T(R) = \rho_T(V) - \beta_T U_T$.
Note that we can define the additional term in the  above equation 
in the Lorentz covariant way by rewriting $U_T^2$ in the tensor field
including six independent components 
though this modification does not change the result. 
  
In order to examine the vacuum effects, here, 
we choose $\beta_T$ to set the vacuum contribution to be zero at $\rho_B=0$
and compare those results with those without the vacuum effect.

The lattice QCD calculation have shown shows that 
the negative magnetic  susceptibility at the zero temperature limit
is zero \cite{LQCD14} or negative \cite{LQCD12}
The magnetic susceptibility is proportional to the spin-susceptibility,
and hence our choice is reasonable for a test calculation.

Then, we take  $\beta_T$ to be
\begin{equation}
\beta_T =  \frac{N_d}{4 \pi^2} \Lambda^2
\left[ 2 F_1 \left( \frac{M_0^2}{\Lambda^2} \right)
+ F_2 \left( \frac{M_0^2}{\Lambda^2} \right) \right] ,
\end{equation}
where $M_0$ is the quark dynamical mass at $\rho_B = 0$.

\begin{figure}[htb]
\begin{center}
\vspace{0.5cm}
{\includegraphics[scale=0.6]{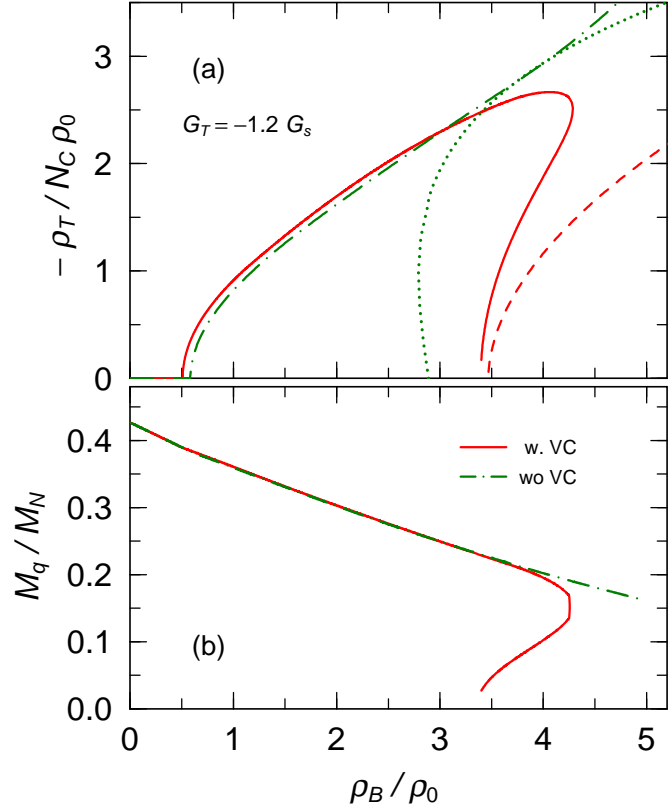}}
\caption{\small
The tensor density normalized by the normal nuclear matter density (a) 
and the dynamical quark mass (b) with PM1 and  $G_T = - G_s$. 
The dot-dashed and solid lines 
represent the results with and without the vacuum contribution (VC), 
respectively.
In the upper panel,
the dotted and -dashed lines indicate the tensor density  
with and without VC
in the chiral-restored phases, respectively. \label{TSV1} }
\end{center}
\end{figure}

In Fig.~\ref{TSV1} we show the tensor density normalized by normal
nuclear matter density $\rho_T/(N_c \rho_0)$ (a) and the dynamical
quark mass $M_q$ with PM1 and $G_T=-1.2G_s$.
The solid and dot-dashed lines represent the results 
without and with the vacuum contribution, respectively.

In the density region, $\rho_B \lesssim \rho_c$. 
the results with the vacuum contribution are almost the same as 
those without the vacuum contribution.
In the density region, $\rho_B \gtrsim \rho_c$, however, 
 the SP ratio is larger than that without the vacuum contribution,
and the SP-I phase  survives when the vacuum contribution is included.
The vacuum polarization has a role to keep the quark
mass finite in the SP phase in high density region.

These large vacuum contributions is considered to come from  the second
term of Eq.~(\ref{RhTLM}), which is proportional to
$\ln(\Lambda^2/M_q^2)$, and becomes larger as the quark dynamical mass 
decrease. 
This contribution cannot be removed by a usual renormalization process
because  a related counter term must be proportional to $M_q^2 U_T^2$, 
which becomes smaller with the decrease of $M_q$.

\end{document}